\documentclass[nofootinbib,prd,12pt,superscriptaddress]{revtex4}%
\usepackage{amsmath}
\usepackage{amsfonts}
\usepackage{amssymb}
\usepackage{graphicx}%
\usepackage{amsmath,amssymb}
\usepackage{mathrsfs}
\usepackage{graphicx}
\usepackage{color}
\usepackage{subfigure}
\usepackage{fancyhdr}
\usepackage{multirow}
\usepackage{float}
\usepackage{epsfig}
\usepackage{amsfonts}
\usepackage{bm}

\def\be{\begin{equation}}
\def\ee{\end{equation}}
\def\ba{\begin{eqnarray}}
\def\ea{\end{eqnarray}}

\begin{document}

\title{Gravity’s Rainbow Effects on Higher Curvature Modification \\of $R^{2}$ inflation}

\author{Jureeporn Yuennan} 
\email{jureeporn\_yue@nstru.ac.th}
\affiliation{Faculty of Science and Technology, Nakhon Si Thammarat Rajabhat University, Nakhon Si Thammarat, 80280, Thailand}

\author{Phongpichit Channuie}
\email{phongpichit.ch@mail.wu.ac.th}
\affiliation{School of Science, Walailak University, Nakhon Si Thammarat, 80160, Thailand}
\affiliation{College of Graduate Studies, Walailak University, Nakhon Si Thammarat, 80160, Thailand}

\author{Davood Momeni}
\affiliation{Northern Virginia Community College, 8333 Little River Turnpike, Annandale, VA 22003, USA}
\affiliation{Centre for Space Research, North-West University, Potchefstroom 2520, South Africa}

\date{\today}

\begin{abstract}

In this work, we study several extensions of the higher curvature modification of $R^{2}$ inflation in the context of gravity's rainbow. We modify the $(R+R^{2})$ model by adding an $f_{1}R^3$-term, an $f_{2}R^4$-term, and an $f_{3}R^{3/2}$-term to the original model. We calculate the inflationary observables and confront them using the latest observational bounds from Planck 2018 data. We assume the rainbow function of the form $\tilde{f}=1+\left(\frac{H}{M}\right)^{\lambda }$ with $\lambda$ being a rainbow parameter and $M$ a mass-dimensional parameter. We demonstrate that the power spectrum of curvature perturbation relies on the dimensionless coefficient $f_{i},\,i=1,2,3$, a rainbow parameter $\lambda$ and a ratio $H/M$. Likewise, the scalar spectral index $n_s$ is affected by both $f_{i}$ and the rainbow parameter. Moreover, the tensor-to-scalar ratio $r$ is solely determined by the rainbow parameter. Interestingly, by ensuring that $n_s$ aligns with the Planck collaboration's findings at the $1\sigma$ confidence level, the tensor-to-scalar ratio could reach up to $r\sim 0.01$, which is possibly measurable for detection in forthcoming Stage IV CMB ground experiments and is certainly feasible for future dedicated space missions.

\end{abstract}

\maketitle


\section{Introduction}
Quantum gravity aims to reconcile two seemingly conflicting principles in physics: Einstein's general theory of relativity, which explains gravity on cosmic scales, and quantum mechanics, which governs physics on the atomic scale and below. All quantum gravity theories, such as loop quantum gravity, string theory, and non-commutative geometry, share a common feature: the existence of a fundamental length scale, often equated with the Planck length. This fundamental length scale requires adjustments to the special theory of relativity, resulting in a modified principle known as doubly special relativity (DSR), which incorporates both the speed of light and the Planck length as fundamental constants. One prominent model that emerges from the DSR framework is gravity's rainbow, which suggests that the curvature of space-time near massive objects like black holes causes light of different wavelengths to travel at varying speeds, creating a "rainbow effect." In this paradigm, the energy of a test particle influences the geometry of space-time, necessitating a family of metrics, termed a rainbow of metrics, rather than a single metric, to describe space-time across all energy scales. These metrics are parameterized by the ratio of the energy of the test particle $(\varepsilon)$ to the Planck energy $(\varepsilon_{p})$. 

At energy scales nearing the Planck scale within different quantum gravity theories, it's anticipated that the traditional dispersion relation will undergo modifications. These alterations are especially significant in light of research such as that examining deformations observed by experiments like the Cherenkov array \cite{Cherenkov}. Initially proposed by Magueijo and Smolin in Ref.\cite{Magueijo:2002xx}, the modification of the dispersion relation replaces the standard expression $\varepsilon^{2}-p^{2}=m^{2}$ with a new form $\varepsilon^{2}{\tilde f}^{2}(\varepsilon) - p^{2}{\tilde g}^{2}(\varepsilon)=m^{2}$, where ${\tilde f}(\varepsilon)$ and ${\tilde g}(\varepsilon)$ are termed rainbow functions. These functions are required to meet certain criteria, particularly approaching unity as energy decreases to an infrared (IR) limit, indicated by ${\tilde f}(\varepsilon/M) \rightarrow 1$ and ${\tilde g}(\varepsilon/M)\rightarrow 1$, where $M$ denotes the energy scale at which quantum gravitational effects start to become notable. 

The influence of rainbow functions has been explored in various cosmological contexts, see, e.g., a correspondence between Horava-Lifshitz gravity and gravity’s rainbow \cite{Garattini:2014rwa}, including black hole physics \cite{Feng:2016zsj,Hendi:2016hbe,Feng:2017gms,Hendi:2018sbe,Panahiyan:2018fpb,Dehghani:2018qvn,Upadhyay:2018vfu,Dehghani:2018svw,Hendi:2017pld,Hendi:2016njy,Hendi:2015cra,Hendi:2015hja,Ali:2015iba,Ali:2014zea,Ali:2014qra,Ali:2014yea,Ali:2014cpa,EslamPanah:2018ums}. Gravity's rainbow has also been examined in alternative gravitational frameworks, including Gauss-Bonnet gravity \cite{Hendi:2016tiy}, massive gravity \cite{Hendi:2017vgo,Heydarzade:2017rpb}, and $f(R)$ gravity \cite{Hendi:2016oxk}. The implications of gravity's rainbow on early universe physics have been widely recognized. In Ref.\cite{Chatrabhuti:2015mws}, the impacts of rainbow functions on the Starobinsky model of $f(R)$ gravity has been examined \cite{Chatrabhuti:2015mws}. More recently, investigations into the deformed Starobinsky model have also been conducted within the framework of gravity's rainbow \cite{Channuie:2019kus}. The analysis also encompassed the work of \cite{Waeming:2020rir} as well as an asymptotic safety theory on inflation within the framework of gravity’s rainbow \cite{Channuie:2024sba}.

In this work, we study several extensions of the higher curvature modification of $R^{2}$ inflation in the context of gravity's rainbow. 
This paper is organized as follows: In Section (\ref{sec2}), we establish a framework for $f(R)$ theory within the context of gravity's rainbow, building upon existing literature \cite{Sotiriou:2008rp,DeFelice:2010aj}. We adopt various $f(R)$ models: (1) $f_{R^{3}}(R)=R+\tfrac{R^2}{6 M^2}+\tfrac{f_1 R^3}{36 M^4}$, (2) $f_{R^{4}}(R)=R+\tfrac{R^2}{6 M^2}+\tfrac{f_{2} R^4}{24 M^6}$ and (3) $f_{R^{3/2}}(R)=R+\tfrac{R^2}{6 M^2}+\tfrac{f_{3} R^{3/2}}{M}$. In section (\ref{sec3}), we take a short recap on cosmological linear perturbations within the framework of gravity's rainbow. Here, we analyze the spectral index of scalar perturbations and the tensor-to-scalar ratio of the models. Additionally, we compare our predicted results with data from Planck 2018 in this section. Finally, we summarize our findings in the last section.

\section{Setup}\label{sec2}
Einstein's theory of gravity, while fundamental, faces contemporary challenges such as dark matter, dark energy, and cosmic inflation. As a result, there is growing interest in modifying general relativity, especially in the early universe where departures from Einstein's theory may become apparent at high curvatures. One approach involves replacing the Einstein-Hilbert term with a function of the Ricci scalar, leading to $f(R)$ theories. Prior research, including pioneering studies on $f(R)$ and alternative gravity theories \cite{Nojiri:2010wj, Nojiri:2017ncd}, has paved the way for further exploration in this area. Here, we embark on our investigation starting with the conventional 4-dimensional action in $f(R)$ gravity, which accounts for matter fields \cite{Sotiriou:2008rp, DeFelice:2010aj},
\ba
S = \frac{1}{2\kappa^2}\int d^4x \sqrt{-g}f(R) + \int d^4x \sqrt{-g}{\cal L}_{M}(g_{\mu\nu},\Psi_{M})\,,
\label{act}
\ea
Here, we define $\kappa^{2}=8\pi G=8\pi/M^{2}{p}$, where $g$ represents the determinant of the metric $g{\mu\nu}$, and the Lagrangian density ${\cal L}{M}$ depends on both $g{\mu\nu}$ and the matter fields $\Psi_{M}$. The field equation can be obtained directly by varying the action (\ref{act}) with respect to $g_{\mu\nu}$ \cite{Sotiriou:2008rp,DeFelice:2010aj},
\ba
F(R)R_{\mu\nu}(g) - \frac{1}{2}f(R)g_{\mu\nu}-\nabla_{\mu}\nabla_{\nu}F(R)+g_{\mu\nu}\Box F(R) = \kappa^{2}T^{(M)}_{\mu\nu}\,,
\label{eom}
\ea
In this context, $F(R)=\partial f(R)/\partial R$, and the operator $\Box$ is defined as $\Box\equiv (1/\sqrt{-g})\partial_{\mu}(\sqrt{-g}g^{\mu\nu}\partial_{\nu})$. The energy-momentum tensor of the matter fields is essentially defined as $T^{(M)}{\mu\nu}=(-2/\sqrt{-g})\delta(\sqrt{-g}{\cal L}{M})/\delta g^{\mu\nu}$, satisfying the continuity equation $\nabla^{\mu}T^{(M)}{\mu\nu}=0$. It's notable that within the conventional approach, the energy-momentum tensor of matter takes on the perfect fluid configuration: $T^{(M)}{\mu\nu}={\rm diag}(-\rho,P,P,P)$, where $\rho$ denotes the energy density and $P$ represents the pressure. Additionally, earlier studies have delved into inflation within various modifications of $R^2$ gravity, as evidenced by works such as Ref.\cite{Sebastiani:2013eqa,Myrzakulov:2014hca,Bamba:2014jia,Elizalde:2017mrn,delaCruz-Dombriz:2016bjj}. With this groundwork established, we proceed to derive cosmological solutions to the field equations (\ref{eom}), employing the modified FLRW metric (\ref{modFR}) as
\begin{eqnarray}
{ds}^2(\varepsilon) =-\frac{dt^2}{\tilde{f}^2(\varepsilon)}+\frac{a^2(t)}{\tilde{g}^2(\varepsilon)} \delta_{ij}{dx}^i{dx}^j\,,\label{modFR}
\end{eqnarray}
Here, $a(t)$ represents a scale factor, while ${\tilde f}(\varepsilon)$ and ${\tilde g}(\varepsilon)$ are referred to as rainbow functions. By substituting the aforementioned metric into the field equations (\ref{eom}) and assuming that the stress-energy tensor follows the perfect fluid form, we arrive at:
\begin{eqnarray}
3 \left(FH^2+H\dot{F} \right) -6FH\frac{  \dot{\tilde{g}}}{\tilde{g}} +3 F\frac{ \dot{\tilde{g}}^2}{\tilde{g}^2}+\dot{F}\frac{ \dot{\tilde{f}}}{\tilde{f}}-3 \dot{F}\frac{ \dot{\tilde{g}}}{\tilde{g}}=\frac{F R-f(R)}{2 \tilde{f}^2}+\frac{\kappa ^2 \rho }{\tilde{f}^2}\,,\label{Hdd}
\end{eqnarray}
and
\begin{eqnarray}
&&3 F H^2-3 \dot{F} H+3 F \dot{H}+3 F H\frac{ \dot{\tilde{f}}}{\tilde{f}}-\dot{F}\frac{ \dot{\tilde{f}}}{\tilde{f}}-4 F\frac{ \dot{\tilde{g}}^2}{\tilde{g}^4}+6 F H\frac{ \dot{\tilde{g}}}{\tilde{g}^3}-3 \dot{F}\frac{ \dot{\tilde{g}}}{\tilde{g}^3}+F\frac{ \ddot{\tilde{g}}}{\tilde{g}^3}+F\frac{\dot{\tilde{f}} }{\tilde{f}}\frac{\dot{\tilde{g}}}{\tilde{g}^3}-3 F H^2\frac{1}{\tilde{g}^2}\nonumber\\&&+2 \dot{F} H\frac{1}{\tilde{g}^2}+\frac{\ddot{F}}{\tilde{g}^2}-F \dot{H}\frac{1}{\tilde{g}^2}+6 F\frac{ \dot{\tilde{g}}^2}{\tilde{g}^2}-F H\frac{ \dot{\tilde{f}}}{\tilde{f} }\frac{1}{\tilde{g}^2}+\dot{F}\frac{ \dot{\tilde{f}}}{\tilde{f}}\frac{1}{\tilde{g}^2}-6 F H\frac{ \dot{\tilde{g}}}{\tilde{g}}+3 \dot{F}\frac{ \dot{\tilde{g}}}{\tilde{g}}-3 F\frac{ \ddot{\tilde{g}}}{\tilde{g}}-3 F\frac{\dot{\tilde{f}} }{\tilde{f} }\frac{\dot{\tilde{g}}}{\tilde{g}} \nonumber\\&&-\frac{f(R) \left(\tilde{g}-1\right) \left(\tilde{g}+1\right)}{2 \tilde{f}^2 \tilde{g}^2}=-\frac{\kappa ^2 \left(\rho  \tilde{g}^2+P\right)}{\tilde{f}^2 \tilde{g}^2}\,,\label{ijcom}
\end{eqnarray}
In this context, we define the first and second derivatives with respect to time as ${\dot a}$ and ${\ddot a}$, respectively. In our subsequent analysis, we set $\tilde{g}=1$ and exclusively consider the spatially flat universe. Consequently, the above equations simplify to:
\begin{eqnarray}
3 FH^2+3H\dot{F}+\dot{F}\frac{ \dot{\tilde{f}}}{\tilde{f}}=\frac{F R-f(R)}{2 \tilde{f}^2}+\frac{\kappa ^2 \rho }{\tilde{f}^2}\,,\label{Hdd}
\end{eqnarray}
and
\begin{eqnarray}
&&3 F H^2-3 \dot{F} H+3 F \dot{H}+3 F H\frac{ \dot{\tilde{f}}}{\tilde{f}}-\dot{F}\frac{ \dot{\tilde{f}}}{\tilde{f}}-3 F H^2+2 \dot{F} H+\ddot{F}-F \dot{H}\nonumber\\&&-F H\frac{ \dot{\tilde{f}}}{\tilde{f}}+\dot{F}\frac{ \dot{\tilde{f}}}{\tilde{f}}=-\frac{\kappa ^2 \left(\rho+P\right)}{\tilde{f}^2}\,.\label{ijcom}
\end{eqnarray}
In the analysis below, we choose
\ba
\tilde{f}=1+\left(\frac{H}{M}\right)^{\lambda }\,,
\ea
Considering the rainbow parameter $\lambda > 0$, we set $\tilde{f}=1$ for late times. Assuming the inflationary period where $H^2 \gg M^2$, we approximate the rainbow function as $\tilde{f}\approx\left(\frac{H}{M}\right)^{\lambda }$.

\subsection{With $R^{3}$ term}
It is worth mentioning that the quantum corrections local effective action which is consistent with diffeomorphism invariance and contains up to fourth order derivatives have been written in Refs.\cite{Codello:2015mba,Oikonomou:2022bqb}. Here the standard quantum corrections to a simple action include $R^{2},\,R^{3}$, non-local terms and Gauss-Bonnet (GB) or higher order derivatives of the form of Riemmann and Ricci tensors or combinations of those. In recent years, there has been substantial interest in exploring higher curvature extensions of the $R^2$ model, see e.g., Refs.\cite{Asaka:2015vza,Huang:2013hsb,Motohashi:2014tra,Bamba:2015uma}. Here we consider the first model of $f(R)$ of the form:
\ba
f_{R^{3}}(R)=R+\frac{R^2}{6 M^2}+\frac{f_1 R^3}{36 M^4}\,,\label{fr3}
\ea
Here, $M$ represents a mass-dimensional parameter. To align with current constraints on $(n_{s}, r)$, it was determined in Ref.\cite{Ivanov:2021chn} for pure Einstein gravity that the dimensionless coefficient $f_{1}$ of $R^{3}$ should satisfy $f_{1}\lesssim {\cal O}(10^{-4})$. This implies that any higher curvature correction beyond the $R^2$ term should not significantly alter the background evolution during $N \sim (50-60)$ e-folds from the end of inflation. Utilizing equation (\ref{fr3}), we derive:
\begin{eqnarray}
F_{R^{3}}(R)=f_{R^{3}}'(R)&\equiv& \frac{\partial f_{R^{3}}(R)}{\partial R}= \frac{f_1 R^2}{12 M^4}+\frac{R}{3 M^2}+1\,,\\
f_{R^{3}}''(R)\equiv \frac{\partial^{2} f_{R^{3}}(R)}{\partial R^{2}}&=&\frac{f_1 R}{6 M^4}+\frac{1}{3 M^2}\,,
\end{eqnarray}
The function $f_{R^{3}}(R)$ adheres to the quantum stability condition $f''_{R^{3}}(R)>0$ for $\alpha >0$ and $\beta >0$, ensuring solution stability at high curvatures. Moreover, the condition of classical stability yields:
\begin{eqnarray}
f_{R^{3}}'(R) = \frac{f_1 R^2}{12 M^4}+\frac{R}{3 M^2}+1 >0\,.
\end{eqnarray}
From Eq.(\ref{Hdd}), we find for this model
\begin{eqnarray}
&&\frac{1}{H M}\Bigg(-6 \text{f1} H^2 \left(2 H^6-19 H^4 (\lambda +1) {\dot H}\right) \left(\frac{H}{M}\right)^{4 \lambda }\nonumber\\&&\quad\quad\quad\quad+3 H^4 M^4+(\lambda +1) M^2 {\dot H} \left(18 H^4+H^2 (17 \lambda -3) {\dot H}\right) \left(\frac{H}{M}\right)^{2 \lambda }\Bigg)=0\,,\label{2sft}
\end{eqnarray}
and from (\ref{ijcom})
\begin{eqnarray}
&&\frac{M^4 {\dot R} \left(\sqrt{3} {\tilde f} \sqrt{\frac{R}{{\tilde f}^2}} \left(12 M^4-f_{1} R^2\right)-12 R {\dot {\tilde f}} \left(f_{1} R+2 M^2\right)\right)}{{\tilde f} R}\nonumber\\&&+2 {\dot R}^2 \left(f_{1}^2 R^2+(6 f_{1}+4) M^4+4 f_{1} M^2 R\right)+12 M^4 {\ddot R} \left(f_{1} R+2 M^2\right)=0\,. \label{22sft}
\end{eqnarray}
In this context, we solely focus on an inflationary solution. Thus, we employ the slow-roll approximations. Consequently, terms containing $\ddot{H}$ and higher powers in $\dot{H}$ can be disregarded in this specific regime. It can be shown straightforwardly that Eq. (\ref{2sft}) is reduced to:
\begin{eqnarray}
\dot{H}\simeq \frac{\left(\frac{H}{M}\right)^{-2 \lambda } \left(4 f_1 H^4 \left(\frac{H}{M}\right)^{4 \lambda }-M^4\right)}{2 (\lambda +1) \left(19 f_1 H^2 \left(\frac{H}{M}\right)^{2 \lambda }+3 M^2\right)}\,.\label{HdHumod2}
\end{eqnarray}
Note that when setting $f_{1}=0$, the result converts to that of Ref.\cite{Chatrabhuti:2015mws}:
\ba
\dot{H}\to -\frac{M^2 \left(\frac{H}{M}\right)^{-2 \lambda }}{6 (\lambda +1)}\,.
\ea
During inflation we can assume $H\simeq {\rm constant.}$, and then in this situation we obtain from Eq.(\ref{2sft})
\ba
H &\simeq& H_i+\frac{\left(\frac{H_{i}}{M}\right)^{-2 \lambda } \left(4 f_1 H^4 \left(\frac{H_{i}}{M}\right)^{4 \lambda }-M^4\right)}{2 (\lambda +1) \left(19 f_1 H^2 \left(\frac{H_{i}}{M}\right)^{2 \lambda }+3 M^2\right)}(t-t_{i})\,,\label{Htkmod2}
\ea
and 
\ba
a \simeq a_i \exp \Bigg\{H_i(t-t_i)+\frac{\left(\frac{H_{i}}{M}\right)^{-2 \lambda } \left(4 f_1 H^4 \left(\frac{H_{i}}{M}\right)^{4 \lambda }-M^4\right)}{2 (\lambda +1) \left(19 f_1 H^2 \left(\frac{H_{i}}{M}\right)^{2 \lambda }+3 M^2\right)}\frac{(t-t_{i})^{2}}{2}\Bigg\},
\ea
In this equation, $H_i$ and $a_i$ represent the Hubble parameter and the scale factor at the onset of inflation ($t=t_i$) respectively. The slow-roll parameter $\epsilon_1$ is defined as $\epsilon_1 \equiv -\dot{H}/H^2$, which in this instance can be approximated as:
\begin{eqnarray}
\epsilon_{1}\equiv -\frac{\dot{H}}{H^{2}}\simeq -\frac{\left(\frac{H}{M}\right)^{-2 \lambda } \left(4 f_1 H^4 \left(\frac{H}{M}\right)^{4 \lambda }-M^4\right)}{2 (\lambda +1)H^{2} \left(19 f_1 H^2 \left(\frac{H}{M}\right)^{2 \lambda }+3 M^2\right)}.\label{epumod2}
\end{eqnarray}
We can check that $\epsilon_{1}$ is less than unity during inflation ($H \gg M$) and we find when setting $f_{1}=0$ that the above expression reduces to $\epsilon_1 \simeq\frac{H^{-2 (\lambda +1)} M^{2 \lambda +2}}{6 (\lambda +1)}$. One can simply determine the time when inflation ends ($t=t_f$) by solving $\epsilon(t_f) \simeq 1$ to obtain
\ba
t_f &\simeq& t_i -\frac{2 (\lambda +1) H_i \left(\frac{H_i}{M}\right)^{2 \lambda } \left(19 f_{1} H_i^2 \left(\frac{H_i}{M}\right){}^{2 \lambda }+3 M^2\right)}{M^4-4 f_{1} H_i^4 \left(\frac{H_i}{M}\right)^{4 \lambda }}\, . \label{tfmod1}
\ea
The number of e-foldings from $t_i$ to $t_f$ is then given by
\ba
N &\equiv& \int^{t_f}_{t_i} Hdt \nonumber\\&\simeq& H_i(t-t_i)+\frac{\left(\frac{H_{i}}{M}\right)^{-2 \lambda } \left(4 f_1 H^4 \left(\frac{H_{i}}{M}\right)^{4 \lambda }-M^4\right)}{2 (\lambda +1) \left(19 f_1 H^2 \left(\frac{H_{i}}{M}\right)^{2 \lambda }+3 M^2\right)}\frac{(t-t_i)^{2}}{2}\nonumber\\&\simeq& \frac{1}{2\epsilon_{1}(t_{i})}\, .
\ea
Note that when $f_{1}=0$, the result is the same as that of the Starobinsky model. In the following section, we examine the spectra of perturbations within the framework of gravity's rainbow theory. We then confront the results predicted by our models with Planck 2018 data.

\subsection{With $R^{4}$ term}
Here we consider the second model of $f(R)$ of the form:
\ba
f_{R^{4}}(R)=R+\frac{R^2}{6 M^2}+\frac{f_{2} R^4}{24 M^6}\,,\label{fr3}
\ea
with $M$ being a mass-dimensional parameter. We noticed that higher-order terms including $R^{4}$ are extensively discussed in non-supersymmetric \cite{Saidov:2010wx, Huang:2013hsb, Sebastiani:2013eqa, Codello:2014sua, Ben-Dayan:2014isa, Artymowski:2014gea, Rinaldi:2014gua, Broy:2014xwa, Artymowski:2015mva} as well as supergravity theories \cite{Cecotti:1987sa, Farakos:2013cqa, Ferrara:2013kca, Ketov:2013dfa, Ozkan:2014cua, Diamandis:2015xra}. To be compatible with the present constraints on $(n_{s}, r)$, it was found in Ref.\cite{Ivanov:2021chn} for pure Einstein gravity that the dimensionless coefficient $f_{2}$ of $R^{4}$ must be $f_{2}\lesssim {\cal O}(10^{-7})$. From the equation (\ref{Hdd}), we obtain
\begin{eqnarray}
F_{R^{4}}(R)=f_{R^{4}}'(R)&\equiv& \frac{\partial f_{R^{4}}(R)}{\partial R}= \frac{f_{2} R^3}{6 M^6}+\frac{R}{3 M^2}+1 \,,\\
f_{R^{4}}''(R)\equiv \frac{\partial^{2} f_{R^{4}}(R)}{\partial R^{2}}&=&\frac{f_{2} R^2}{2 M^6}+\frac{1}{3 M^2}\,,
\end{eqnarray}
The function $f_{R^{4}}(R)$ obeys the quantum stability condition $f''_{R^{4}}(R)>0$ for $\alpha >0$ and $\beta >0$. This ensures the stability of the solution at high curvature. Additionally, the condition of classical stability leads to
\begin{eqnarray}
f_{R^{4}}'(R) = \frac{f_{2} R^3}{6 M^6}+\frac{R}{3 M^2}+1>0\,.
\end{eqnarray}
From Eq.(\ref{Hdd}), we find for this model
\begin{eqnarray}
&&-\frac{432 f_2 H^9 \left(\frac{H}{M}\right)^{6 \lambda }}{M}+\frac{3888 f_2 H^7 \lambda  \left(\frac{H}{M}\right)^{6 \lambda } H'(t)}{M}+\frac{3888 f_2 H^7 \left(\frac{H}{M}\right)^{6 \lambda } \dot{H}}{M}\nonumber\\&&+3 H^3 M^5+18 H^3 M^3 \left(\frac{H}{M}\right)^{2 \lambda } \dot{H}+18 H^3 \lambda  M^3 \left(\frac{H}{M}\right)^{2 \lambda}\dot{H}=0\,,\label{2sft}
\end{eqnarray}
and from (\ref{ijcom})
\begin{eqnarray}
&&-\frac{2 M^6 {\dot R} \left(\sqrt{3} {\tilde f} \sqrt{\frac{R}{{\tilde f}^2}} \left(f_{2} R^3-3 M^6\right)+{\dot{\tilde f}} \left(9 f_{2} R^3+6 M^4 R\right)\right)}{{\tilde f} R}\nonumber\\&&+\left({\dot R}\right)^2 \left(9 f_{2}^2 R^4+36 f_{2} M^6 R+12 f_{2} M^4 R^2+4 M^8\right)+6 M^6 {\ddot R} \left(3 f_{2} R^2+2 M^4\right)=0\,. \label{22sft}
\end{eqnarray}
Here we are only interested in an inflationary solution. Therefore we invoke the slow-roll approximations. Hence the terms containing $\ddot{H}$ and higher power in $\dot{H}$ can be neglected in this particular regime. It is rather straightforward to show that the Eq.(\ref{2sft}) is reduced to
\ba
\dot{H}\simeq \frac{\left(\frac{H}{M}\right)^{-2 \lambda } \left(144 f_2 H^6 \left(\frac{H}{M}\right)^{6 \lambda }-M^6\right)}{6 (\lambda +1) \left(216 f_2 H^4 \left(\frac{H}{M}\right)^{4 \lambda }+M^4\right)}.\label{HdHumod2}
\ea
Note that when setting $f_{2}=0$, the result converts to that of Ref.\cite{Chatrabhuti:2015mws}:
\ba
\dot{H}\to -\frac{M^2 \left(\frac{H}{M}\right)^{-2 \lambda }}{6 (\lambda +1)}\,.
\ea
During inflation we can assume $H\simeq {\rm constant.}$, and then in this situation we obtain from Eq.(\ref{2sft})
\ba
H &\simeq& H_i+\frac{\left(\frac{H_{i}}{M}\right)^{-2 \lambda } \left(144 f_2 H_{i}^6 \left(\frac{H_i}{M}\right)^{6 \lambda }-M^6\right)}{6 (\lambda +1) \left(216 f_2 H_{i}^4 \left(\frac{H_i}{M}\right)^{4 \lambda }+M^4\right)}(t-t_i)\,,\label{Htkmod2}
\ea
and 
\ba
a \simeq a_i \exp \Bigg\{H_i(t-t_i)+\frac{\left(\frac{H_{i}}{M}\right)^{-2 \lambda } \left(144 f_2 H_{i}^6 \left(\frac{H_i}{M}\right)^{6 \lambda }-M^6\right)}{6 (\lambda +1) \left(216 f_2 H_{i}^4 \left(\frac{H_i}{M}\right)^{4 \lambda }+M^4\right)}\frac{(t-t_i)^{2}}{2}\Bigg\},
\ea
where $H_i$ and $a_i$ are respectively the Hubble parameter and the scale factor at the onset of inflation ($t=t_i$). The slow-roll parameter $\epsilon_1$ is defined by $\epsilon_1 \equiv -\dot{H}/H^2$ which in this case can be estimated as
\begin{eqnarray}
\epsilon_{1}\equiv -\frac{\dot{H}}{H^{2}}\simeq \frac{\left(\frac{H}{M}\right)^{-2 \lambda } \left(144 f_2 H^6 \left(\frac{H}{M}\right)^{6 \lambda }-M^6\right)}{6 (\lambda +1)H^{2} \left(216 f_2 H^4 \left(\frac{H}{M}\right)^{4 \lambda }+M^4\right)}.\label{epumod2}
\end{eqnarray}
We can check that $\epsilon_{1}$ is less than unity during inflation ($H \gg M$) and we find when setting $f_{2}=0$ that the above expression reduces to $\epsilon_1 \simeq\frac{H^{-2 (\lambda +1)} M^{2 \lambda +2}}{6 (\lambda +1)}$. One can simply determine the time when inflation ends ($t=t_f$) by solving $\epsilon(t_f) \simeq 1$ to obtain
\ba
t_f &\simeq& t_i + \frac{6 (\lambda +1)H_i \left(\frac{H_i}{M}\right)^{2 \lambda } \left(216 f_2 H_i^4 \left(\frac{H_i}{M}\right)^{4 \lambda }+M^4\right)}{M^6-144 f_2 H_i^6 \left(\frac{H_i}{M}\right)^{6 \lambda}}\, . \label{tfmod1}
\ea
The number of e-foldings from $t_i$ to $t_f$ is then given by
\ba
N &\equiv& \int^{t_f}_{t_i} Hdt \nonumber\\&\simeq& H_i(t-t_i)+\frac{\left(\frac{H_{i}}{M}\right)^{-2 \lambda } \left(144 f_2 H_{i}^6 \left(\frac{H_i}{M}\right)^{6 \lambda }-M^6\right)}{6 (\lambda +1) \left(216 f_2 H_{i}^4 \left(\frac{H_i}{M}\right)^{4 \lambda }+M^4\right)}\frac{(t-t_i)^{2}}{2}\nonumber\\&\simeq& \frac{1}{2\epsilon_{1}(t_{i})}\, .
\ea
Note that when $f_{2}=0$, the result is the same as that of the Starobinsky model. In the subsequent section, we study the perturbation spectra within the framework of gravity's rainbow theory. Following this examination, we compare our model's predictions with the data reported by Planck 2018 dataset.

\subsection{With $R^{3/2}$ term}
We consider the last model of $f(R)$ of the form:
\ba
f_{R^{3/2}}(R)=R+\frac{R^2}{6 M^2}+\frac{f_{3} R^{3/2}}{M}\,,\label{fr33}
\ea
with $M$ being a mass-dimensional parameter. Note that $R^{3/2}$ corrections are intriguing for the exploration of modifications to gravity suggesting that it might provide a theoretical basis for modified newtonian dynamics (MOND) \cite{Bernal:2011qz}, possibly linking empirical observations with a comprehensive framework and offering insights into galactic gravitational interactions. From the equation (\ref{fr33}), we obtain
\begin{eqnarray}
F_{R^{3/2}}(R)=f_{R^{3/2}}'(R)&\equiv& \frac{\partial f_{R^{3/2}}(R)}{\partial R}= \frac{3 f_3 \sqrt{R}}{2 M}+\frac{R}{3 M^2}+1\,,\\
f_{R^{3/2}}''(R)\equiv \frac{\partial^{2} f_{R^{3/2}}(R)}{\partial R^{2}}&=&\frac{3 f_3}{4 M \sqrt{R}}+\frac{1}{3 M^2}\,,
\end{eqnarray}
The function $f_{R^{3/2}}(R)$ obeys the quantum stability condition $f''_{R^{3/2}}(R)>0$ for $f_3 >0$. This ensures the stability of the solution at high curvature. Additionally, the condition of classical stability leads to
\begin{eqnarray}
f_{R^{3/2}}'(R) = \frac{3 f_3 \sqrt{R}}{2 M}+\frac{R}{3 M^2}+1 >0\,.
\end{eqnarray}
From Eq.(\ref{Hdd}), we find for this model
\begin{eqnarray}
&&12 \sqrt{6} f_3 H^6 M \left(\frac{H}{M}\right)^{2 \lambda }+6 H^4 \Bigg((\lambda +1) \left(\frac{H}{M}\right)^{2 \lambda } {\dot H} \Bigg(5 \sqrt{6} f_3 M+12 \times\nonumber\\&&\times\sqrt{\left(\frac{H}{M}\right)^{2 \lambda } \left(2 H^2+(\lambda +1) {\dot H}\right)}\Bigg)+2 M^2 \sqrt{\left(\frac{H}{M}\right)^{2 \lambda } \left(2 H^2+(\lambda +1) {\dot H}\right)}\Bigg)=0\,,\label{2sft}
\end{eqnarray}
and from (\ref{ijcom})
\begin{eqnarray}
&&6 {\dot R} \left(\sqrt{3} {\tilde{f}} M^3 \sqrt{R} \sqrt{\frac{R}{{\tilde{f}}^2}} \left(3 f_{3} \sqrt{R}+4 M\right)-2 M^2 R {\dot{\tilde{f}}} \left(9 f_{3} M+4 \sqrt{R}\right)\right)\nonumber\\&&+f \left({\dot R}\right)^2 \left(81 f_{3}^2 M^2 \sqrt{R}-54 f_{3} M^3+72 f_{3} M R+16 R^{3/2}\right)\nonumber\\&&+12 {\tilde{f}} M^2 R {\ddot R} \left(9 f_{3} M+4 \sqrt{R}\right)=0\,. \label{22sft}
\end{eqnarray}
Here we are only interested in an inflationary solution. Therefore we invoke the slow-roll approximations. Hence the terms containing $\ddot{H}$ and higher power in $\dot{H}$ can be neglected in this particular regime. It is rather straightforward to show that the Eq.(\ref{2sft}) is reduced to
\ba
\dot{H}\simeq -\frac{M^2 \left(\frac{H}{M}\right)^{-2 \lambda } \left(\sqrt{3} f_3 \left(\frac{H}{M}\right)^{\lambda +1}+1\right)}{6 (\lambda +1)}.\label{HdHumod2}
\ea
Note that when setting $f_{3}=0$, the result converts to that of Ref.\cite{Chatrabhuti:2015mws}:
\ba
\dot{H}\to -\frac{M^2 \left(\frac{H}{M}\right)^{-2 \lambda }}{6 (\lambda +1)}\,.
\ea
During inflation we can assume $H\simeq {\rm constant.}$, and then in this situation we obtain from Eq.(\ref{2sft})
\ba
H &\simeq& H_i-\frac{M^2 \left(\frac{H_i}{M}\right)^{-2 \lambda } \left(\sqrt{3} f_3 \left(\frac{H_i}{M}\right)^{\lambda +1}+1\right)}{6 (\lambda +1)}(t-t_i)\,,\label{Htkmod2}
\ea
and 
\ba
a \simeq a_i \exp \Bigg\{H_i(t-t_i)-\frac{M^2 \left(\frac{H}{M}\right)^{-2 \lambda } \left(\sqrt{3} f_3 \left(\frac{H}{M}\right)^{\lambda +1}+1\right)}{6 (\lambda +1)}\frac{(t-t_i)^{2}}{2}\Bigg\},
\ea
where $H_i$ and $a_i$ are respectively the Hubble parameter and the scale factor at the onset of inflation ($t=t_i$). The slow-roll parameter $\epsilon_1$ is defined by $\epsilon_1 \equiv -\dot{H}/H^2$ which in this case can be estimated as
\begin{eqnarray}
\epsilon_{1}\equiv -\frac{\dot{H}}{H^{2}}\simeq \frac{M^2 \left(\frac{H}{M}\right)^{-2 \lambda } \left(\sqrt{3} f_3 \left(\frac{H}{M}\right)^{\lambda +1}+1\right)}{6 (\lambda +1)H^{2}}.\label{epumod2}
\end{eqnarray}
We can check that $\epsilon_{1}$ is less than unity during inflation ($H \gg M$) and we find when setting $f_{3}=0$ that the above expression reduces to $\epsilon_1 \simeq\frac{H^{-2 (\lambda +1)} M^{2 \lambda +2}}{6 (\lambda +1)}$. One can simply determine the time when inflation ends ($t=t_f$) by solving $\epsilon(t_f) \simeq 1$ to obtain
\ba
t_f &\simeq& t_i + \frac{6 (\lambda +1) \left(\frac{H_i}{M}\right)^{2 \lambda +1}}{M \left(\sqrt{3} f_3 \left(\frac{H_i}{M}\right){}^{\lambda +1}+1\right)}\, . \label{tfmod1}
\ea
The number of e-foldings from $t_i$ to $t_f$ is then given by
\ba
N &\equiv& \int^{t_f}_{t_i} Hdt \nonumber\\&\simeq& H_i(t-t_i)-\frac{M^2 \left(\frac{H_i}{M}\right)^{-2 \lambda } \left(\sqrt{3} f_3 \left(\frac{H_i}{M}\right)^{\lambda +1}+1\right)}{6 (\lambda +1)}\frac{(t-t_i)^{2}}{2}\nonumber\\&\simeq& \frac{1}{2\epsilon_{1}(t_{i})}\, .
\ea
In the subsequent section, we analyze perturbation spectra in the context of gravity's rainbow theory. Subsequently, we compare the results predicted by our model with the data from Planck 2018.

\section{Confrontation with Observation}\label{sec3}
In this section, we closely follow the approaches outlined in References \cite{Channuie:2019kus, Waeming:2020rir} for examining cosmological linear perturbations within the context of gravity's rainbow induced by inflation. Detailed derivations of the spectral index of curvature perturbation and the tensor-to-scalar ratio are provided therein, sparing us from reiterating them here. To determine the curvature perturbation spectrum, we begin by defining the slow-roll parameters:
\begin{eqnarray}
\epsilon_1 \equiv -\frac{\dot{H}}{H^2}, \ \ \epsilon_2 \equiv \frac{\dot{F}}{2HF}, \ \ \epsilon_3 \equiv \frac{\dot{E}}{2HE}\ .
\end{eqnarray}\label{a20}
where $E \equiv 3\dot{F}^2/2\kappa^2$. Afterwards, $Q_s$ can be reformulated to yield:
\begin{eqnarray}
Q_s = \frac{E}{FH^2(1+\epsilon_2)^2} \ . \label{a22}
\end{eqnarray}
During the inflationary era, parameters $\epsilon_i$ are assumed to be constant ($\dot{\epsilon}i\simeq0$), and in this work, we take $\tilde{f} =1+(H/M)^{\lambda}$. Therefore, the spectral index can be written in terms of the slow-roll parameters as:
\begin{eqnarray}
n{s} - 1 \simeq -2(\lambda+2)\epsilon_1+2\epsilon_2-2\epsilon_3 \ , \label{a31}
\end{eqnarray}
In the inflationary phase, we've presumed that $|\epsilon_i | \ll 1$. It's worth noting that the spectrum tends towards scale invariance when $|\epsilon_i|$ are significantly below unity, implying $n_{s} \simeq 1$. Following this, the power spectrum of the curvature perturbation adopts the following structure:
\begin{eqnarray}
\mathcal{P}{S} \approx \frac{1}{Q_s}\left(\frac{H}{2\pi}\right)^2\left(\frac{H}{M}\right)^{2\lambda} \ . \label{a32}
\end{eqnarray}
Based on the aforementioned outcome, we can limit the parameters for the $f(R)$ models by utilizing Eq.(\ref{a32}), given that $\mathcal{P}{S}$ encompasses $Q_s$, which depends on $f(R)$. Subsequently, the power spectrum of tensor perturbations $P_T$ following the Hubble radius crossing can be approximated as:
\begin{eqnarray}
\mathcal{P}T \simeq \frac{16}{\pi}\left(\frac{H}{M{p}}\right)^2\frac{1}{F}\left(\frac{H}{M}\right)^{2\lambda} \ . \label{a42}
\end{eqnarray}
Also, the tensor-to-scalar ratio $r$ can be determined by invoking the following definition:
\begin{eqnarray}
r \equiv \frac{\mathcal{P}T}{\mathcal{P}S} \simeq 48\epsilon_2^2 ,. \label{a43}
\end{eqnarray}
In the subsequent section, we examine the perturbation spectra within the framework of different $f(R)$ models in gravity's rainbow theory, posing the observables predicted by our models with the Planck 2018 data. As indicated in References \cite{Channuie:2019kus, Waeming:2020rir}, a connection between $\varepsilon{1}$ and $\varepsilon{1}$ can be quantified:
\begin{eqnarray}
\epsilon_2 \simeq -(1+\lambda)\epsilon_1 . \label{ep1ep2}
\end{eqnarray}
We can verify another relation among slow-roll parameters by considering the definition of $\epsilon_3$:
\begin{eqnarray}
\epsilon_3 \equiv \frac{\dot{E}}{2HE}=\frac{\ddot{F}}{H \dot{F}}.
\end{eqnarray}
To examine the connections between slow-roll parameters, we will examine various forms of $f(R)$ outlined below. We will utilize the most recent Planck 2018 data \cite{Planck:2018jri, Planck:2018vyg} and the latest BICEP/Keck data \cite{BICEP:2021xfz} to obtain well-defined parameters: the scalar spectral index $n_s$ and the tensor-to-scalar ratio $r$, as specified:
\begin{eqnarray}
    &&{\rm Planck\,2018} : \quad n_{s} = 0.9658 \pm 0.0038 , \quad r < 0.072 \,,\nonumber\\&&
    {\rm BICEP/Keck2021} : \quad r < 0.036 .\nonumber
\end{eqnarray}

\subsection{With $R^{3}$ term}
In the following, we will express the power spectrum, the spectral index, $n_{s}$, and the tensor-to-scalar ratio, $r$, in terms of e-folding number $N$. By substituting the expression for $Q_s$ in Eq.(\ref{a22}) into Eq. (\ref{a32}) and using the relation (\ref{ep1ep2}), we then obtain
\ba
\mathcal{P}_{S} \approx\frac{M^2 N^2}{3 \pi  (\lambda +1)^2 M_p^2}\left(1-3 f_1 \left(\frac{H}{M}\right)^{2 \lambda +2}\right)\label{PsN}
\ea
Using parameters of the base $\Lambda$CDM cosmology reported by Planck 2018 for $P_{S}$ at the scale $k=0.05\,{\rm Mpc}^{-1}$, we find from Eq.(\ref{PsN}) that
\ba
\frac{M}{M_{p}}\equiv X \simeq  \Bigg(\frac{2.07\times 10^{-8}(\lambda +1)^2}{N^2 \left(1-3f_{1} \left(\frac{H_{*}}{M}\right)^{2\lambda +2}\right)}\Bigg)^{\tfrac{1}{2}}
\ea
Taking $N= 60,\, \lambda = 0.02,\,f_{1}\sim {\cal O}(10^{-7})$ and $H_{*}/M \in [10^{2},\,10^{3}]$, we find that $X\sim {\cal O}(10^{-6})$. Moreover, if we increase $H_{*}/M$ such that $H_{*}/M>10^{3}$, values of $f_{1}$ must be extremely small, $f_{1}\ll {\cal O}(10^{-7})$. Assuming further that $X\sim {\cal O}(10^{-4})$ and using $H_{*}/M=10^{5},\,N= 60,\, \lambda = 0.02$, we can obtain $f_{1}\sim {\cal O}(10^{-11})$. From Eq.(\ref{a31}) and Eq.(\ref{a43}), we obtain
\ba
n_{s}&\approx&1-\frac{2}{N}-\frac{12 f_{1} (\lambda +1)}{N}\left(\frac{H_{*}}{M}\right)^{2 (\lambda +1)},\\
r&\approx&\frac{12(1+\lambda)^{2}}{N^{2}}\,.
\ea

\begin{figure}[ht!]
    \centering
\includegraphics[width=4.5in,height=4.5in,keepaspectratio=true]{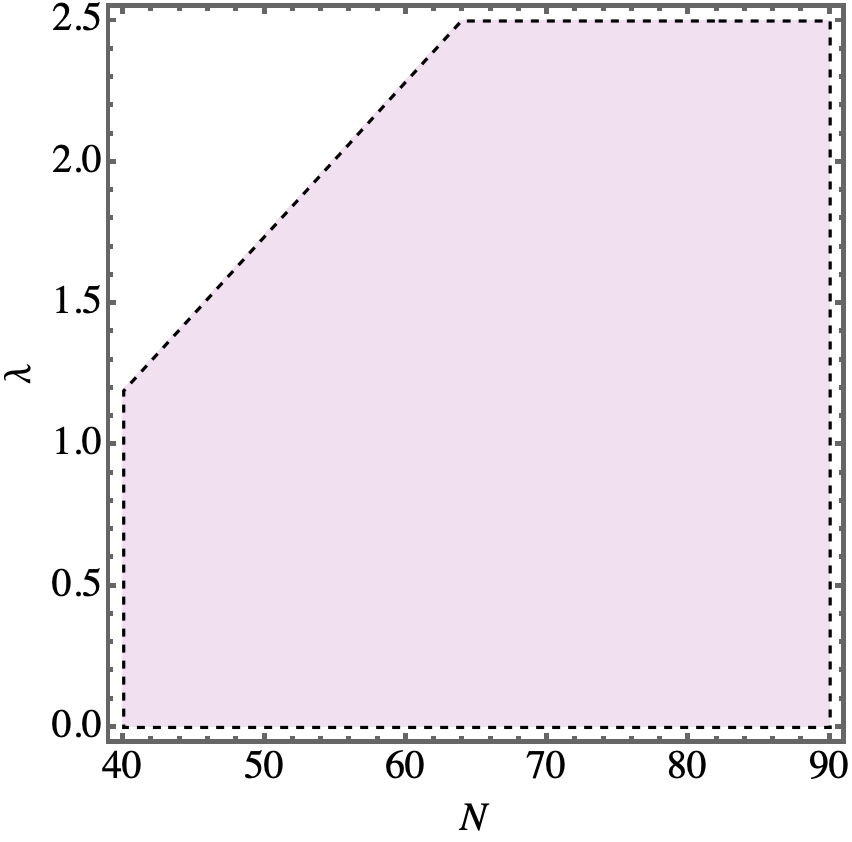}
    \caption{We quantify the upper limit of $\lambda$ for any $N$, with for example $\lambda < 1.74\,(2.23)$ for $N=50\,(60)$. A shaded area represents an inequality given by $\frac{12(1+\lambda)^{2}}{N^{2}}<0.036$.}
    \label{upper}
\end{figure}

Clearly, setting $f_{1}=0$, the results become those of the Starobinsky model of inflation in the context of gravity's rainbow. Moreover, setting both $\lambda=0$ and $f_{1}=0$, the Starobinsky model of inflation in the Einsten gravity can be obtained. We confront the results predicted by our model with Planck 2018 data. In order to be consistent with data, we find that the value of $\lambda$ cannot be arbitrary. Its upper limit can be displayed in Fig.(\ref{upper}) and computed using $r<0.036$ to obtain $\lambda < 1.74\,(2.23)$ for $N=50\,(60)$.

\begin{figure}[ht!]
    \centering
\includegraphics[width=5in,height=5in,keepaspectratio=true]{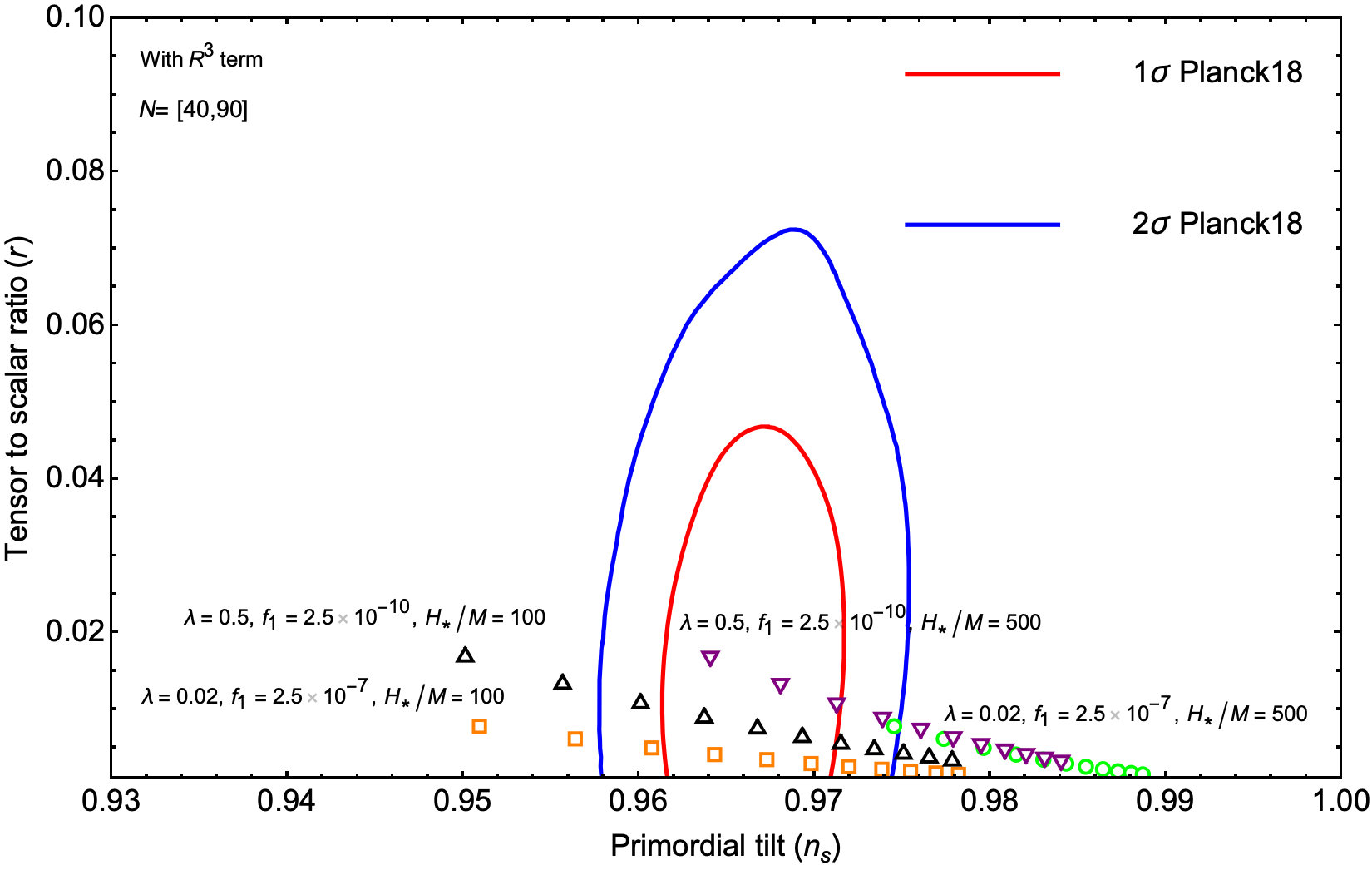}
    \caption{We compare the theoretical predictions of $f_{R^{3}}(R)=R+\frac{R^2}{6 M^2}+\frac{f_1 R^3}{36 M^4}$ in the $(r-n_{s})$ plane for different values of $N$ ranging from $[40,\,90]$ (left to right) assuming a various set of parameters, $\lambda,\,f_{1}$ and $H_{*}/M$.}
    \label{rnss1}
\end{figure}

Requiring $n_s$ to be consistent with the Planck collaboration upper limit, we find that $r$ can be as large as $r\simeq 0.01$ illustrated in Fig.(\ref{rnss1}). More precisely, using $\lambda=0.02,\,f_{1}=2.5\times 10^{-7},\,\tfrac{H_{*}}{M}=100$, we obtain $r\sim 0.005$ for $N\in [55,\,60]$ (Orange). Moreover, requiring $n_s$ to be consistent with the Planck collaboration at $1\,\sigma$\,confidence level (CL.), values of $r$ could be as large as $r\sim 0.01$ when using $\lambda=0.5,\,f_{1}=2.5\times 10^{-7},\,\tfrac{H_{*}}{M}=100$. However, the predictions taking $\lambda=0.02,\,f_{1}=2.5\times 10^{-7},\,\tfrac{H_{*}}{M}=500$ (Green) lie outside the $1\,\sigma$ confidence level.

\subsection{With $R^{4}$ term}
\begin{figure}[ht!]
    \centering
\includegraphics[width=5in,height=5in,keepaspectratio=true]{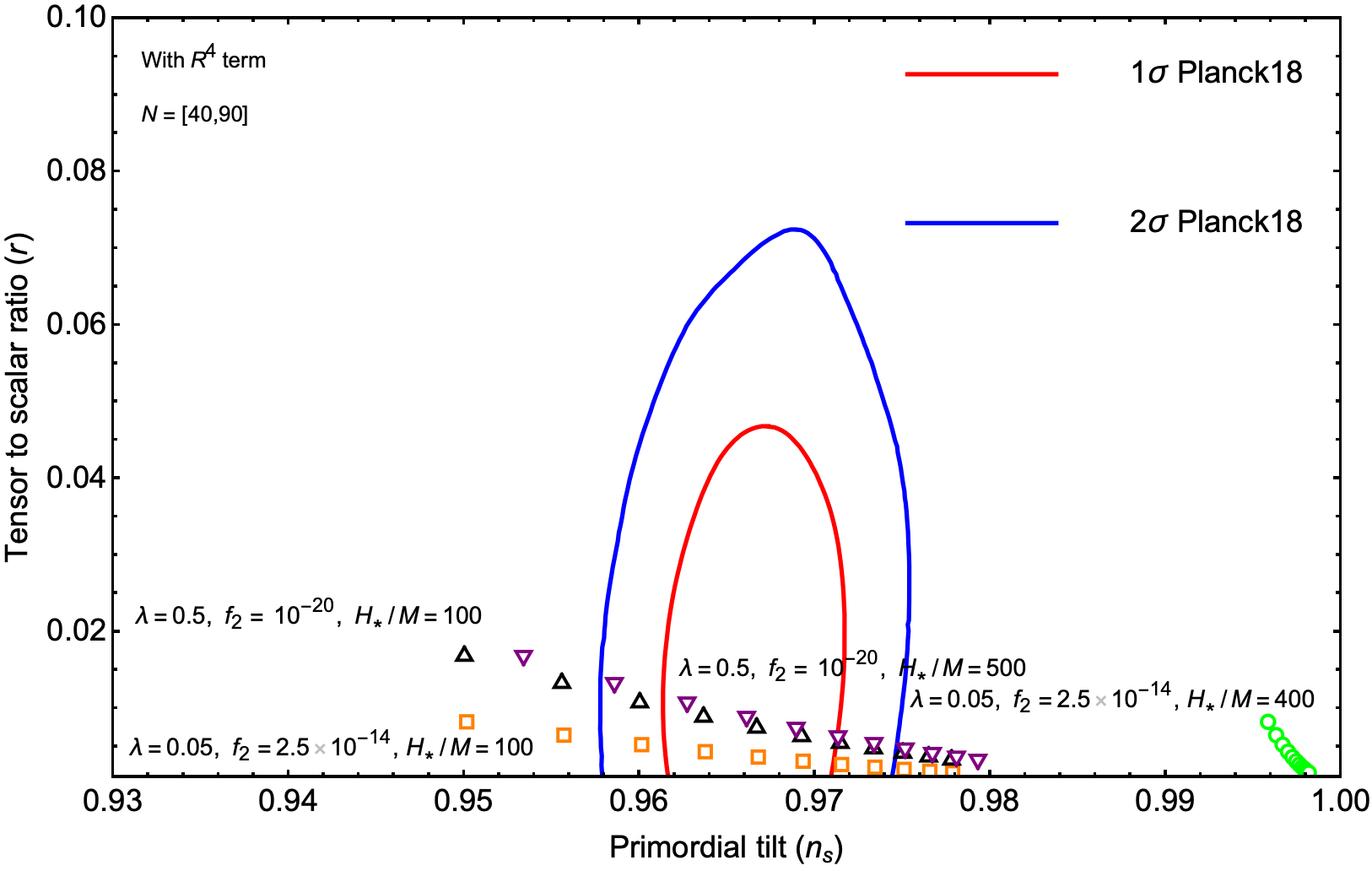}
    \caption{We compare the theoretical predictions of $f_{R^{4}}(R)=R+\frac{R^2}{6 M^2}+\frac{f_{2} R^4}{24 M^6}$ in the $(r-n_{s})$ plane for different values of $N$ ranging from $[40,\,90]$ (left to right) assuming a various set of parameters, $\lambda,\,f_{2}$ and $H_{*}/M$.}
    \label{rnss2}
\end{figure}

In the following, we will express the power spectrum, the spectral index, $n_{s}$, and the tensor-to-scalar ratio, $r$, in terms of e-folding number $N$. By substituting the expression for $Q_s$ in Eq.(\ref{a22}) into Eq. (\ref{a32}) and using the relation (\ref{ep1ep2}), we then obtain
\ba
\mathcal{P}_{S} \approx\frac{M^2 N^2}{3 \pi  (\lambda +1)^2 M_p^2}\left(1-72 f_1 \left(\frac{H}{M}\right)^{4 (\lambda +1)}\right)\label{PsN2}
\ea
Using parameters of the base $\Lambda$CDM cosmology reported by Planck 2018 for $P_{S}$ at the scale $k=0.05\,{\rm Mpc}^{-1}$, we find from Eq.(\ref{PsN2}) that
\ba
\frac{M}{M_{p}}\equiv X \simeq  \Bigg(\frac{2.07\times 10^{-8}(\lambda +1)^2}{N^2 \left(1-72f_{1} \left(\frac{H_{*}}{M}\right)^{4 (\lambda +1)}\right)}\Bigg)^{\tfrac{1}{2}}\,.
\ea
Taking $N= 60,\, \lambda = 0.001,\,f_{2}\sim {\cal O}(10^{-14})$ and $H_{*}/M \in [10^{2},\,10^{3}]$, we find that $X\sim {\cal O}(10^{-6})$. Moreover, if we increase $H_{*}/M$ such that $H_{*}/M>10^{3}$, values of $f_{1}$ must be extremely small, $f_{2}\ll {\cal O}(10^{-14})$. Assuming further that $X\sim {\cal O}(10^{-4})$ and using $H_{*}/M=10^{5},\,N= 60,\, \lambda = 0.01$, we can obtain $f_{2}\sim {\cal O}(10^{-24})$. From Eq.(\ref{a31}) and Eq.(\ref{a43}), we obtain
\ba
n_{s}&\approx&1-\frac{2}{N}-\frac{864 f_2 (\lambda +1)}{N}\left(\frac{H_{*}}{M}\right)^{4 (\lambda +1)},\\
r&\approx&\frac{12(1+\lambda)^{2}}{N^{2}}\,.
\ea
It is evident that when $f_{2}$ is set to zero, the results coincide with those derived from the Starobinsky model of inflation within the framework of gravity's rainbow. Moreover, setting both $\lambda=0$ and $f_{2}=0$, the Starobinsky model of inflation in the Einsten gravity can be obtained. We confront the results predicted by our model with Planck 2018 data. Requiring $n_s$ to be consistent with the Planck collaboration upper limit, we find that $r$ can be as large as $r\simeq 0.01$ illustrated in Fig.(\ref{rnss2}). More precisely, using $\lambda=0.02,\,f_{2}=2.5\times 10^{-14},\,\tfrac{H_{*}}{M}=100$, we obtain $r\sim 0.005$ for $N\in [55,\,60]$ (Orange). Moreover, requiring $n_s$ to be consistent with the Planck collaboration at $1\,\sigma$\,CL, values of $r$ could be higher, e.g., $r\sim 0.01$ when using $\lambda=0.5,\,f_{2}=2.5\times 10^{-20},\,\tfrac{H_{*}}{M}=100$ for $N\in [55,\,60]$. However, the predictions taking $\lambda=0.05,\,f_{2}=2.5\times 10^{-14},\,\tfrac{H_{*}}{M}=400$ (Green) lie far outside the $2\,\sigma$ confidence level.

\subsection{With $R^{3/2}$ term}
\begin{figure}[ht!]
    \centering
\includegraphics[width=5in,height=5in,keepaspectratio=true]{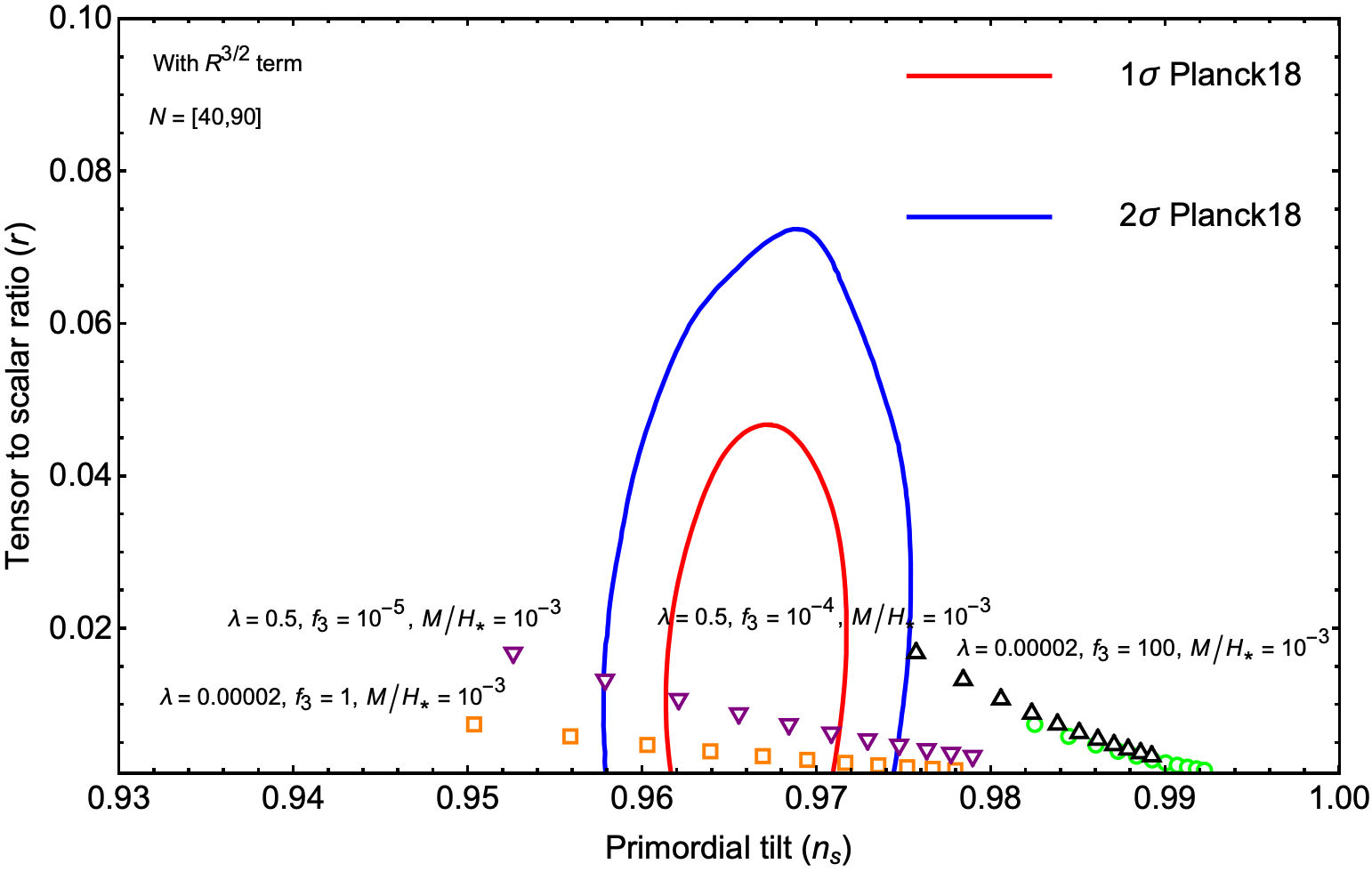}
    \caption{We compare the theoretical predictions of $f_{R^{3/2}}(R)=R+\frac{R^2}{6 M^2}+\frac{f_{3} R^{3/2}}{M}$ in the $(r-n_{s})$ plane for different values of $N$ ranging from $[40,\,90]$ (left to right) assuming a various set of parameters, $\lambda,\,f_{3}$ and $H_{*}/M$.}
    \label{rnss32}
\end{figure}

In the following, we will express the power spectrum, the spectral index, $n_{s}$, and the tensor-to-scalar ratio, $r$, in terms of e-folding number $N$. By substituting the expression for $Q_s$ in Eq.(\ref{a22}) into Eq. (\ref{a32}) and using the relation (\ref{ep1ep2}), we then obtain
\ba
\mathcal{P}_{S} \approx\frac{M^2 N^2}{3\pi (\lambda +1)^2 M_p^2 \left(1+\frac{3}{4}\sqrt{3} f_3 \left(\frac{M}{H}\right)^{\lambda +1} \right)}\label{PsN3}
\ea
Using parameters of the base $\Lambda$CDM cosmology reported by Planck 2018 for $P_{S}$ at the scale $k=0.05\,{\rm Mpc}^{-1}$, we find from Eq.(\ref{PsN3}) that
\ba
\frac{M}{M_{p}}\equiv X \simeq  \Bigg(\frac{2.69 \times 10^{-8}(\lambda +1)^2}{N^2}\Big(f_3 \left(\frac{M}{H_{*}}\right)^{\lambda +1}+0.7698\Big)\Bigg)^{\tfrac{1}{2}}
\ea
Taking $N= 60,\, \lambda = 0.02,\,f_{3}\in [0.1-1000],\,M/H_{*} = {\cal O}(10^{-4})$, we find that $X\sim {\cal O}(10^{-6})$. From Eq.(\ref{a31}) and Eq.(\ref{a43}), we obtain
\ba
n_{s}&\approx&1-\frac{2}{N}+\frac{3 \sqrt{3} f_{2} \lambda  \left(\frac{H_{*}}{M}\right)^{-\lambda -1}}{8 N},\\
r&\approx&\frac{12(1+\lambda)^{2}}{N^{2}}\,.
\ea
Upon setting $f_3$ to zero, the outcomes coincide with those derived from the Starobinsky model of inflation within the framework of gravity's rainbow. Moreover, setting both $\lambda=0$ and $f_{1}=0$, the Starobinsky model of inflation in the Einsten gravity can be obtained. We confront the results predicted by our model with Planck 2018 data. Requiring $n_s$ to be consistent with the Planck collaboration upper limit, we find that $r$ can be as large as $r\simeq 0.01$ illustrated in Fig.(\ref{rnss32}). More precisely, using $\lambda=2.0\times 10^{-5},\,f_{3}=1.0,\,\tfrac{M}{H_{*}}=10^{-3}$, we obtain $r\sim 0.005$ for $N\in [55,\,60]$ (Orange). Moreover, values of $r$ could be as large as $r\sim 0.01$ when using $\lambda=0.5,\,f_{3}=10^{-5},\,\tfrac{M}{H_{*}}=10^{-3}$. However, the predictions taking $\lambda=2.0\times 10^{-5},\,f_{3}=100,\,\tfrac{M}{H_{*}}=10^{-3}$ (Green) lie far outside the $2\,\sigma$ confidence level.

\section{Concluding Remarks}
In this work, we have considered several extensions of the higher curvature modification of $R^{2}$ inflation in the context of gravity's rainbow. We have modified the $(R+R^{2})$ model by adding an $R^3$-term, an $R^4$-term, and an $R^{3/2}$-term. We have calculated the inflationary observables and confront them using the latest observational data. The COBE normalization and the Planck constraint on the scalar spectrum have been used to constrain the predictions.

We have demonstrated that the power spectrum of curvature perturbation relies on the dimensionless coefficient $f_{i},\,i=1,2,3$, a rainbow parameter $\lambda$ and a ratio $H/M$. Likewise, the scalar spectral index $n_s$ is affected by both $f_{i}$ and the rainbow parameter. Moreover, the tensor-to-scalar ratio $r$ is exclusively determined by the rainbow parameter. Interestingly, by ensuring that $n_s$ aligns with the Planck collaboration's findings at the $1\sigma$ confidence level, the tensor-to-scalar ratio could reach up to $r\sim 0.01$. This value is possibly measurable for detection in forthcoming Stage IV CMB ground experiments and is certainly feasible for future dedicated space missions. 

Additionally, the impact of the reheating era in gravity's rainbow theories with $R^{2}$ and higher-order corrections is worth investigating. A quick approach to this is to follow the analysis provided in Refs.\cite{DeFelice:2010aj}. Specifically, to study particle production during reheating, we consider a scalar field $\chi$ with mass $m_{\chi}$ or in the massless case \cite{Motohashi:2012tt}. We also introduce a non-minimal coupling term $(1/2)\xi R\chi^{2}$ between the field $\chi$ and the Ricci scalar $R$ \cite{Birrell:1982ix}. Next, we redefine the coordinates such that $dt = \tilde{f}\, dt_{1}$ and $a = \tilde{g}\, \tilde{a}$. From this point, the procedure follows standard methods, with the results in Ref.\cite{DeFelice:2010aj} being modified by taking $t \rightarrow t_{1}$ and $a \rightarrow \tilde{a}$. At this stage, the reheating temperature can be estimated similarly to that in Ref.\cite{DeFelice:2010aj}. The difference lies in the mass scale $M$, which emerges from different underlying descriptions. However, a detailed analysis of the reheating dynamics in the present work needs to be thoroughly examined, and we also leave this important issue for further investigation. 


\end{document}